\documentclass[iop]{emulateapj}

\def\kms{km~s$^{-1}$}
\usepackage{graphicx}

\def\dlambda{$\lambda\lambda$}

\slugcomment{Submitted to The Astrophysical Journal on August 6, 2014}

\shorttitle{The Broad-lined Type Ic SN\,2012ap}
\shortauthors{D.\ Milisavljevic et al.}

\newenvironment{my_itemize}{
\begin{itemize}
	\setlength{\itemsep}{1pt}
	\setlength{\parskip}{2pt}
	\setlength{\parsep}{1pt}}{\end{itemize}
}

\begin{document}

\def\cfa{1}
\def\dartmouth{2}
\def\ari{3}
\def\mp{4}
\def\inafpad{5}
\def\kyoto{6}
\def\uu{7}
\def\gsfc{8}
\def\berk{9}
\def\ut{10}
\def\salt{11}
\def\stsci{12}
\def\hasc{13}
\def\subaru{14}
\def\lco{15}
\def\aarhus{16}
\def\usz{17}

\title{The Broad-Lined Type I\lowercase{c} SN\,2012\lowercase{ap} and
  the Nature of Relativistic Supernovae\\ Lacking a Gamma-ray Burst
  Detection}

\author{D.~Milisavljevic\altaffilmark{\cfa,$\dagger$},
        R.~Margutti\altaffilmark{\cfa},
        J.~T.~Parrent\altaffilmark{\cfa},
        A.~M.~Soderberg\altaffilmark{\cfa},                
        R.~A.~Fesen\altaffilmark{\dartmouth},
        P.~Mazzali\altaffilmark{\ari,\mp,\inafpad},
        K.~Maeda\altaffilmark{\kyoto,\uu},
        N.~E.~Sanders\altaffilmark{\cfa},
        S.~B.~Cenko\altaffilmark{\gsfc,\berk},
        J.~M.~Silverman\altaffilmark{\ut},
        A.~V.~Filippenko\altaffilmark{\berk},       
        A.~Kamble\altaffilmark{\cfa},
        S.~Chakraborti\altaffilmark{\cfa},    
        M.~R.~Drout\altaffilmark{\cfa},
        R.~P.~Kirshner\altaffilmark{\cfa},
        T.~E.~Pickering\altaffilmark{\salt,\stsci},
        K.~Kawabata\altaffilmark{\hasc},
        T.~Hattori\altaffilmark{\subaru},
        E.~Y.~Hsiao\altaffilmark{\lco,\aarhus},
        M.~D.~Stritzinger\altaffilmark{\aarhus},
        G.~H.~Marion\altaffilmark{\ut},
        J.~Vinko\altaffilmark{\ut,\usz}, and
        J.~C.~Wheeler\altaffilmark{\ut}       
}

\altaffiltext{\cfa}{Harvard-Smithsonian Center for Astrophysics, 60
  Garden St., Cambridge, MA 02138}
\altaffiltext{\dartmouth}{Department of Physics \& Astronomy, Dartmouth
                 College, 6127 Wilder Lab, Hanover, NH 03755, USA}
\altaffiltext{\ari}{Astrophysics Research Institute, Liverpool John Moores University,
Liverpool L3 5RF, United Kingdom}
\altaffiltext{\mp}{Max-Planck-Institut f\"ur Astrophysik,
  Karl-Schwarzschild-Strasse 1, 85748 Garching, Germany}
\altaffiltext{\inafpad}{INAF - Osservatorio Astronomico di Padova,
  Vicolo dell'Osservatorio 5, I-35122, Padova, Italy}
\altaffiltext{\kyoto}{Department of Astronomy, Kyoto University
Kitashirakawa-Oiwake-cho, Sakyo-ku, Kyoto 606-8502, Japan}
\altaffiltext{\uu}{Kavli Institute for the Physics and Mathematics
of the Universe (WPI), Todai Institutes for Advanced Study,
University of Tokyo, 5-1-5 Kashiwanoha, Kashiwa, Chiba 277-8583, Japan}
\altaffiltext{\gsfc}{Astrophysics Science Division, NASA Goddard Space
  Flight Center, Mail Code 661, Greenbelt, MD 20771, USA}
\altaffiltext{\berk}{ Department of Astronomy, University of
  California, Berkeley, CA 94720-3411, USA}
\altaffiltext{\ut}{University of Texas at Austin, 1 University Station
  C1400, Austin, TX, 78712-0259, USA}
\altaffiltext{\salt}{Southern African Large Telescope, PO Box 9,
Observatory 7935, Cape Town, South Africa}
\altaffiltext{\stsci}{Space Telescope Science Institute, 3700 San
  Martin Drive, Baltimore, Maryland 21218, USA}
\altaffiltext{\hasc}{Hiroshima Astrophysical Science Center, Hiroshima University, Higashi-Hiroshima, Hiroshima 739-8526, Japan}
\altaffiltext{\subaru}{Subaru Telescope, National Astronomical Observatory of Japan, Hilo, HI 96720, USA}
\altaffiltext{\lco}{Carnegie Observatories,
  Las Campanas Observatory, Colina El Pino, Casilla 601, Chile}
\altaffiltext{\aarhus}{Department of Physics and Astronomy, Aarhus University, Ny Munkegade, DK-8000 Aarhus C, Denmark}
\altaffiltext{\usz}{Department of Optics and Quantum Electronics,
University of Szeged, Dom ter 9, 6720, Szeged, Hungary}
\altaffiltext{$\dagger$}{email: dmilisav@cfa.harvard.edu}

\begin{abstract}

  We present ultraviolet, optical, and near-infrared observations of
  SN\,2012ap, a broad-lined Type Ic supernova in the galaxy NGC 1729
  that produced a relativistic and rapidly decelerating outflow
  without a gamma-ray burst signature. Photometry and spectroscopy
  follow the flux evolution from $-13$ to $+272$ days past the
  $B$-band maximum of $-17.4 \pm 0.5$ mag. The spectra are dominated
  by \ion{Fe}{2}, \ion{O}{1}, and \ion{Ca}{2} absorption lines at
  ejecta velocities of $v \approx$ 20,000 \kms\ that change slowly
  over time. Other spectral absorption lines are consistent with
  contributions from photospheric \ion{He}{1}, and hydrogen may also
  be present at higher velocities ($v \ga$ 27,000 \kms). We use these
  observations to estimate explosion properties and derive a total
  ejecta mass of $\sim 2.7 \;{\rm M}_{\odot}$, a kinetic energy of
  $\sim 1.0\times10^{52}\;\rm{erg}$, and a $^{56}$Ni mass of $0.1-0.2
  \;{\rm M}{_\odot}$. Nebular spectra ($t > 200$\,d) exhibit an
  asymmetric double-peaked [\ion{O}{1}] $\lambda\lambda$6300, 6364
  emission profile that we associate with absorption in the supernova
  interior, although toroidal ejecta geometry is an alternative
  explanation.  SN\,2012ap joins SN\,2009bb as another exceptional
  supernova that shows evidence for a central engine (e.g., black-hole
  accretion or magnetar) capable of launching a non-negligible portion
  of ejecta to relativistic velocities without a coincident gamma-ray
  burst detection. Defining attributes of their progenitor systems may
  be related to notable properties including above-average
  environmental metallicities of $Z \ga {\rm Z}_{\odot}$, moderate to
  high levels of host-galaxy extinction $(E(B-V) > 0.4$ mag),
  detection of high-velocity helium at early epochs, and a high
  relative flux ratio of [\ion{Ca}{2}]/[\ion{O}{1}] $> 1$ at nebular
  epochs. These events support the notion that jet activity at various
  energy scales may be present in a wide range of supernovae.

\end{abstract}

\section{Introduction}

The spectral features of core-collapse supernovae (SN) provide a basis
of classification that reflects properties of their progenitor stars
and explosion dynamics
\citep{Minkowski41,Shklovskii60,Kirshner73,Oke74}. By standard
definition, Type Ib supernovae lack conspicuous absorptions
attributable to hydrogen, and Type Ic supernovae lack conspicuous
absorptions attributable to hydrogen and helium
\citep{Filippenko97,Matheson01,Turatto03-Classification,Modjaz14}. These
two subgroups, however, may have many deviant cases 
\citep[e.g.,][]{Branch06,Parrent07,James10}, and a
possible continuum between them is sometimes acknowledged by using the
designation Type Ibc (hereafter SN\,Ibc).

SN\,Ibc are thought to originate from stars that have been largely
stripped of their outer envelopes \citep{Wheeler87,Clocchiatti97}, via
radiative winds \citep{Woosley93} or various forms of binary
interaction \citep{Podsiadlowski92,Nomoto95}. No secure direct
identification has yet been made of a SN\,Ibc progenitor system
(\citealt{vanDyk03,Smartt09,Eldridge13}; although see \citealt{Cao13},
\citealt{Bersten14}, and \citealt{Fremling14}).

\begin{figure}[htp!]
\centering
\includegraphics[width=\linewidth]{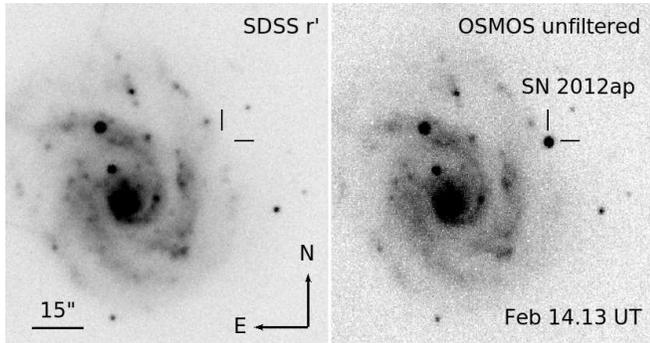}

\caption{Images of SN\,2012ap and its host galaxy NGC\,1729. Left:
  Pre-explosion SDSS $r'$-band image of NGC 1729 with the location of
  SN\,2012ap marked. Right: Unfiltered image obtained with the 2.4\,m
  Hiltner telescope using the OSMOS instrument and MDM4k detector.}

\label{fig:fchart}
\end{figure}

Broad-lined Type Ic supernovae (SN\,Ic-bl) are a subset of SN\,Ibc that
show exceptionally high expansion velocities in their bulk ejecta
reaching $\sim 0.1\;\rm c$. Generally, SN\,Ic-bl are associated with
large kinetic energies (several $10^{52}$ erg) approximately 10 times
those of normal SN\,Ibc, and ejected masses of several M$_{\odot}$, of
which $\sim 0.5\; {\rm M}_{\odot}$ is $^{56}$Ni
\citep{Mazzali08-Progenitor}. However, the handful of SN\,Ic-bl known
are diverse and can vary considerably in these explosion properties
\citep{Nomoto07}. This diversity has been underscored by recent
examples such as SN\,2007bg \citep{Young10}, SN 2007ru \citep{Sahu09},
SN\,2009nz \citep{Berger11}, SN 2010ay \citep{Sanders12-10ay},
PTF10qts \citep{Walker14}, SN\,2010ah \citep{Corsi11,Mazzali13}, and
PTF12gzk \citep{Ben-Ami12}.

A crucial and revealing aspect of SN\,Ic-bl is that they can accompany
long-duration gamma-ray bursts (GRBs). The coincidence of nearby
events including SN\,1998bw with GRB980425 \citep{Galama98} and
SN\,2003dh with GRB030329 \citep{Stanek03,Matheson03} has established
that all well-observed GRB-SN are SN\,Ic-bl. However, as demonstrated
by objects such as SN\,2002ap \citep{Mazzali02,Berger02}, the converse
is not true: namely, not all SN\,Ic-bl are associated with GRBs.

It is an open question as to why some SN\,Ic-bl are associated with
GRBs and others are not.  Radio observations seem to rule out the
possibility that all SN\,Ic-bl without a GRB detection are off-axis
GRBs \citep{Soderberg06GRB}. Thus, certain properties of the
progenitor systems and explosion dynamics of these SN must dictate why
some explosions are not sufficient to generate a GRB.  Notable
properties include the following. (1) SN\,Ic-bl not associated with
GRBs tend to have smaller values of ejecta mass, explosion energy, and
luminosity as compared to the GRB-SN \citep{Nomoto07} --- although
exceptions do exist, such as SN\,2010bh associated with GRB\,100316D
\citep{Chornock10,Olivares12,Bufano12}. (2) The relative rates of GRB
and SN\,Ic-bl are comparable ($10^{-6}$ to $10^{-5}\;\rm yr^{-1}$ per
galaxy) and support the notion that they originate from the same
population \citep{Podsiadlowski04}. (3) As with GRBs, SN\,Ic-bl
preferentially occur in regions of high star-formation rates and/or
very young stellar populations having subsolar metallicity
environments when compared to normal SN\,Ibc
\citep{Kelly12,Sanders12}.

Discovery of SN\,2009bb provided strong evidence of a continuum
between GRB-SN and SN\,Ic-bl. SN\,2009bb was a SN\,Ic-bl that exhibited
radio properties consistent with a non-negligible portion of its
ejecta moving at relativistic speeds as observed in GRBs, yet was
subenergetic by a factor of $\sim 100$ and did not have a GRB
detection \citep{Soderberg10,Pignata11,Chakraborti11}. Because the
bulk explosion parameters of SN\,2009bb could not account for the
copious energy coupled to relativistic ejecta, it was concluded that a
central engine (e.g., black hole accretion or magnetar) must have
driven part of the explosion.

A second and more recent example of this type of event is
SN\,2012ap. As in SN\,2009bb, SN\,2012ap was found to have
relativistic outflow but without an observed GRB
\citep{Chakraborti14}. Moreover, neither object showed evidence for
luminous X-ray emission at late times ($t > 10$\,d), which sets them
apart from subenergetic GRBs \citep{Margutti14}. These shared
properties led \citet{Margutti14} to conclude that this
distinct class of objects may represent the weakest engine-driven
explosions, where the central engine is unable to power a successful
jet breakout.

Here we report on ultraviolet, optical, and near-infrared observations
of SN\,2012ap from $-13$ to $+272$ days past $B$-band maximum. In \S
\ref{sec:discovery} we discuss the discovery and classification of
SN\,2012ap. Section \ref{sec:obs} presents the data, a portion of
which have already been published by \citet{Milisavljevic14}
(hereafter M14). These data are then used in \S \ref{sec:results} to 
examine the flux evolution of the SN, reconstruct its bolometric light 
curve, and derive explosion parameters.  In \S \ref{sec:discussion} we 
discuss the implications our results and analyses have for potential
progenitor systems of SN\,2009bb and SN\,2012ap. Section 
\ref{sec:conclusions} concludes with a review of the properties of 
relativistic SN\,Ic-bl without a GRB detection and speculates on the 
extent to which jet activity at various energy scales may be occurring 
in a wide range of SN.

\section{Discovery and Classification}
\label{sec:discovery}

SN\,2012ap was first detected by the Lick Observatory Supernova Search
\citep[LOSS;][]{Filippenko01} with the 0.76\,m Katzman Automatic Imaging
Telescope (KAIT)
at coordinates $\alpha$(J2000) = 05$^h$00$^m$13$\fs$72 and
$\delta$(J2000) = $-03\degr$20$\arcmin$51$\farcs$2 in NGC\,1729 on Feb
10.23 (UT dates are used throughout this paper) \citep{Jewett12}.  In
Figure~\ref{fig:fchart}, pre- and post-explosion images are shown,
highlighting the location of the SN with respect to NGC\,1729. The SN
is located some 7.1\,kpc in projection from the nucleus of the host
galaxy along the outer periphery of a spiral arm.

\citet{Xu12} obtained optical spectra of SN\,2012ap on February 11 and
12 with the Chinese Gao-Mei-Gu telescope and classified it as a
SN\,Ibc at early phases. They noted a close similarity with the
SN\,Ib\,2008D two weeks before maximum light. \citet{Milisavljevic12}
reported on spectra obtained Feb 21.8 showing similarities with the
broad-lined SN\,1998bw \citep{Patat01}, SN\,2002ap \citep{Foley03},
and the transitional SN\,2004aw \citep{Taubenberger06} approximately
1--2 weeks after maximum light. These later spectra were in general
agreement with the earlier observations reported by Xu et al., but
they no longer showed a strong likeness to SN\,2008D.

\begin{figure}[!htp]
\centering
\includegraphics[width=0.95\linewidth]{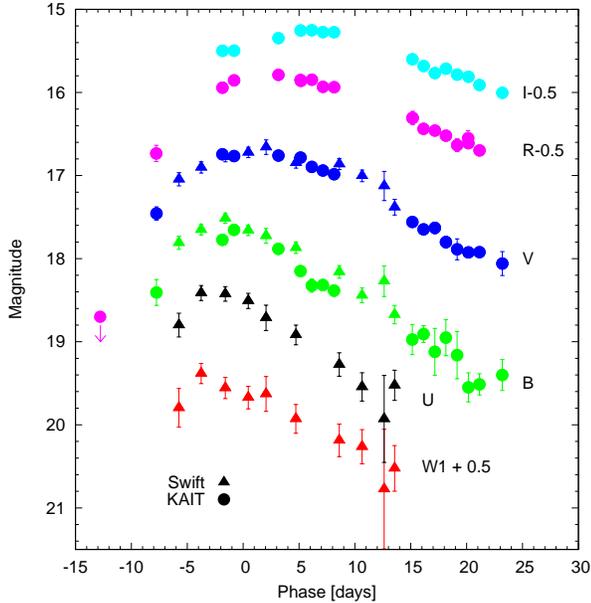}

\caption{KAIT and \emph{Swift} UV/optical photometry of
  SN\,2012ap. Phase is with respect to $B$-band maximum on Feb.\ 18.2
  UT.}

\label{fig:photometry}
\end{figure}

\section{Observations}
\label{sec:obs}

\subsection{Distance and Reddening}

The distance to NGC\,1729 estimated by Tully-Fisher measurements is 
43\,Mpc, corresponding to a distance modulus of $\mu = 33.17 \pm 0.48$\,mag
\citep{Springob09}. Conspicuous \ion{Na}{1}~D absorption at the rest
wavelength of the host galaxy in our spectra (see \S
\ref{sec:obs:spectra}) and a low apparent brightness at maximum light
(\S \ref{sec:obs:phot}) both suggest moderate extinction toward
the supernova. The foreground Galactic extinction is $E(B-V)_{\rm
  Galactic}=0.045$ mag \citep{Schlafly11}. An estimate of the host
extinction was made using the equivalent width (EW) of the
\ion{Na}{1}~D line. Following \citet{Turatto03}, EW(\ion{Na}{1})
$\approx 1.2$\,\AA\ yields estimates of $E(B-V)$ between 0.182 and
0.572 mag. Using the same measurement and following instead
\citet{Poznanski12}, the estimate is $0.36 \pm 0.07$ mag. The mean of
$\sim 0.4$ mag has been used in this paper. Combining the Galactic
extinction with the inferred host extinction, a total extinction of
$E(B-V)_{\rm total}=0.45$ mag has been adopted.

Conspicuous narrow absorption features associated with diffuse
interstellar bands (DIBs) at rest wavelengths of 4428\,\AA, 5780\,\AA,
and 6283\,\AA\ are observed in the optical spectra of SN\,2012ap, and they
change in EW strength between epochs of observations (see M14 for
details). In some settings, DIBs have been used to infer the amount of
foreground extinction because their EW can be linearly proportional to
the amount of foreground reddening
\citep{Herbig95,Friedman11}. However, the unusually strong DIB
absorptions observed in SN\,2012ap are well outside the reliable
limits of these relationships and cannot be used to calibrate the
extinction.

\begin{figure}[htp!]
\centering
\includegraphics[width=\linewidth]{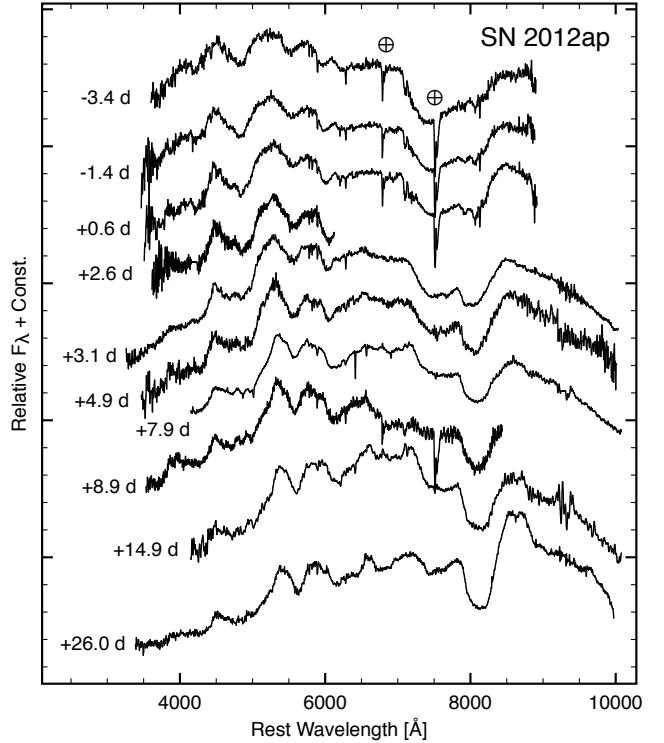}

\caption{Optical spectra of SN\,2012ap. The spectra have been
  corrected for a redshift of 0.0121. The ``$\oplus$'' symbol shows
  regions of the spectra that in some cases are contaminated by
  night-sky O$2$ absorption bands.}

\label{fig:specallepochs}
\end{figure}

\begin{figure}[htp!]
\centering
\includegraphics[width=\linewidth]{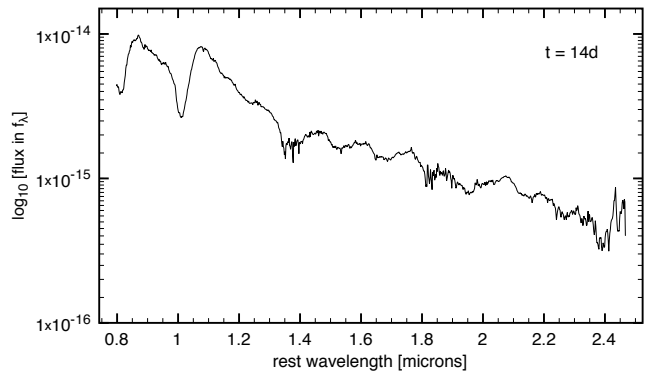}

\caption{Near-infrared spectrum of SN\,2012ap obtained on 2012 March 03
  with the FIRE spectrograph on the 6.5\,m Magellan Baade telescope.}

\label{fig:nirspec}
\end{figure} 

\subsection{UV/Optical Photometry}
\label{sec:obs:phot}

Optical photometry was obtained with KAIT, and both optical and
ultraviolet (UV) photometry was taken with the \emph{Swift} spacecraft
\citep{Gehrels04} using the UVOT instrument \citep{Roming05}. The
observed magnitudes are presented in Tables~\ref{tab:kaitphot} and
\ref{tab:swiftphot}, respectively. KAIT photometry was calibrated using 
Sloan Digital Sky Survey (SDSS) point sources in the field, and photometric
transformations were made from \citet{Jordi06} to put SDSS photometry into
the $BVRI$ system. We analyzed the \emph{Swift}-UVOT photometric data
following the prescriptions of \cite{Brown09}. A $3''$ aperture was
used to maximize the signal-to-noise ratio (S/N).  Unreported are
observations in the $uvw2$ and $uvm2$ filters where the SN was not
detected with a significant S/N. \emph{Swift}-UVOT photometry is
based on the photometric system described by \citet{Breeveld11}. In
this system, the \emph{Swift} $b$ and $v$ filters are roughly
equivalent to the standard Johnson/Kron-Cousins $B$ and $V$ filters
(see \citealt{Poole08} for details), although the difference
introduces a small offset in reported magnitudes between the
\emph{Swift} and KAIT photometry. Figure~\ref{fig:photometry} shows
the combined \emph{Swift}-UVOT and KAIT light curves.

\subsection{Optical and Near-Infrared Spectra}
\label{sec:obs:spectra}

\begin{figure}
\centering
\includegraphics[width=\linewidth]{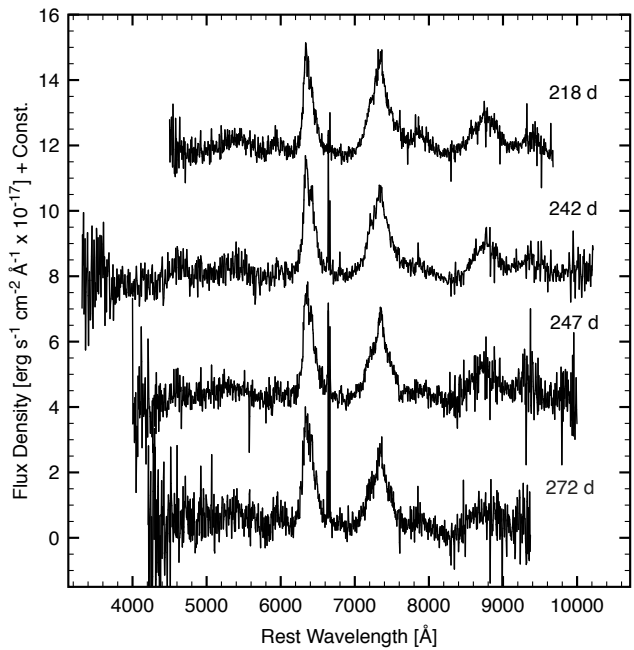}

\caption{Late-time optical spectra of SN\,2012ap during nebular epochs
  when emission is dominated by forbidden transitions.}

\label{fig:latetimespec}
\end{figure}

Fourteen epochs of long-slit optical spectra of SN\,2012ap were
obtained from a variety of telescopes and instruments. A single
near-infrared spectrum was also obtained.  Early-time spectra are
shown in Figure~\ref{fig:specallepochs} (optical) and
Figure~\ref{fig:nirspec} (near-infrared), and late-time spectra are
shown in Figure~\ref{fig:latetimespec}.  Table~\ref{tab:spectra} lists
the details of the observations.

Many observations were made with the 10\,m Southern African Large
Telescope (SALT) at South African Astronomical Observatory using the
Robert Stobie Spectrograph (RSS; \citealt{Burgh03}). Additional
supporting observations came from the 9.2\,m Hobby-Eberly Telescope
(HET) using the Marcario Low-Resolution Spectrograph (LRS;
\citealt{Hill98}), the 6.5\,m MMT telescope using the Blue Channel
instrument \citep{Schmidt89}, the 8.2\,m Subaru Telescope using the
Faint Object Camera and Spectrograph (FOCAS; \citealt{Kashikawa02}),
the 6.5\,m Magellan Baade Telescope using the Inamori Magellan Areal
Camera and Spectrograph (IMACS; \citealt{Bigelow98}), the 10\,m Keck
telescopes using the Low Resolution Imaging Spectrometer (LRIS;
\citealt{Oke95}) and the DEep Imaging Multi-Object Spectrograph
(DEIMOS; \citealt{Faber03}), and the Shane 3\,m telescope using the Kast
double spectrograph \citep{Miller93}. The single near-infrared observation
was obtained with the Magellan 6.5\,m Baade telescope using the
FoldedPort Infrared Echellette (FIRE; \citealt{Simcoe08}).

Reduction of all optical spectra followed standard procedures using
the IRAF/PyRAF software. SALT data were first processed by the
PySALT\footnote{http://pysalt.salt.ac.za/} pipeline
\citep{Crawford10}. Wavelength calibrations were made with arc lamps
and verified with the night-sky lines. Relative flux calibrations were
made with observations of spectrophotometric standard stars from
\citet{Oke90} and \citet{Hamuy92,Hamuy94}. Gaps between CCD chips have
been interpolated in instances when dithering between exposures was
not possible, and cosmetic defects have been cleaned manually. In
cases when a spectrophotometric standard star could be observed at the
time of observation and at comparable airmass, telluric features have
been corrected. Near-infrared data were reduced following standard
procedures \citep{Hsiao13} using a custom-developed IDL
pipeline (FIREHOSE).

Spectra have been corrected for a redshift of $z = 0.0121$ measured from
narrow \ion{H}{2} region lines of [\ion{O}{3}] \dlambda 4959, 5007,
H$\alpha$, and [\ion{N}{2}] \dlambda 6548, 6583 observed near the
location of the supernova. This value is in agreement with a
previous measurement reported by \citet{Theureau98} using radio
\ion{H}{1} lines from the host NGC\,1729.

\begin{figure}
\centering
\includegraphics[width=\linewidth]{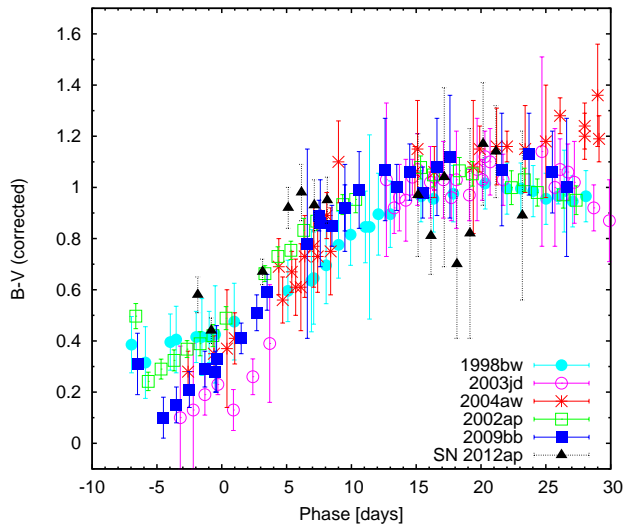}

\caption{Color evolution of SN\,2012ap compared to that of various SN\,Ic and
  Ic-bl. References for data on individual supernovae and adopted
  $E(B-V)$ values are as follows: SN\,1998bw (0.0645 mag;
  \citealt{Clocchiatti11}), SN\,2003jd (0.14 mag;
  \citealt{Valenti08a}), SN\,2004aw (0.37 mag;
  \citealt{Taubenberger06}), SN\,2002ap (0.079 mag;
  \citealt{Foley03}), SN\,2009bb (0.58 mag; \citealt{Pignata11}),
  SN\,2012ap (0.45 mag; this paper).}

\label{fig:b-v}
\end{figure}

\section{Results}
\label{sec:results}

\subsection{Properties of the UV/Optical Light Curves}
\label{sec:lc}

Table~\ref{tab:lcprops} shows the properties of our light curves as
determined with low-order polynomial fits. The peak in the $B$-band
corresponds to an absolute magnitude of $M_B = -17.4 \pm 0.5$. This
value is relatively faint compared to the majority of SN\,Ic-bl
\citep{Drout11,Bianco14}, being well below that of SN 1998bw ($M_B =
-18.8$ mag; \citealt{Patat01}) but above that of SN\,2002ap ($M_B =
-16.3$ mag; \citealt{Foley03}).

A KAIT image of the region around SN\,2012ap was taken Feb 5.21 prior
to detection with limiting $R$-band magnitude of 18.7. This
nondetection sets a constraint on the explosion date, with some
uncertainty caused by the steepness of the unobserved light curve as it
rises and a possible offset owing to a dark phase in the early stages of
the explosion \citep{Nakar12}. We used the light curve of SN\,2009bb
as a template and scaled its peak and width to follow that of
SN\,2012ap. From this we estimate the explosion date to be Feb. 5,
with an uncertainty of $\sim 2$\,d, and derive a rise time to
peak maximum $B$-band light of 13\,d based on the $B$-band maximum on
Feb. 18.2.

\begin{figure}[htp!]
\centering
\includegraphics[width=\linewidth]{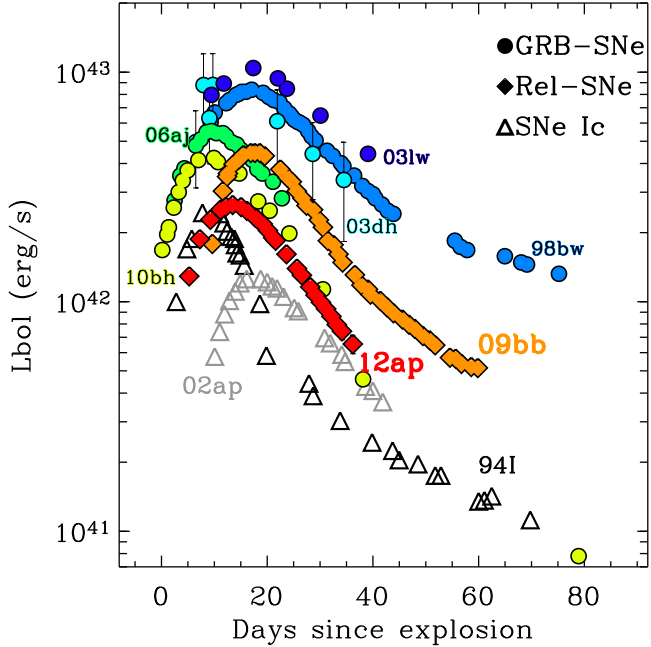}

\caption{Bolometric light curve of SN\,2012ap. Data for all objects
  are obtained from \citet{Olivares12}, except for SN\,2003dh, which is
  from \citet{Pian06}.}

\label{fig:lbol}
\end{figure}

The $B-V$ color evolution of SN\,2012ap is illustrated in
Figure~\ref{fig:b-v}. Also shown are the color evolutions of other 
SN\,Ic and Ic-bl. Within the uncertainties of our photometry, the color
evolution of SN\,2012ap is broadly consistent with those observed in
previous SN\,Ibc. We also find consistency in the $V-R$ colors of
SN\,2012ap with other SN\,Ibc at similar epochs presented by
\citet{Drout11}. The general agreement demonstrates that the adopted
$E(B-V)$ value is reasonable.

The extinction-corrected \emph{Swift}-UVOT and KAIT photometry was
used to create a quasi-bolometric light curve
($L_{\rm{bol}}^{\rm{quasi}}$) of SN\,2012ap.  The total UV+{\it BVRI}
flux was determined by summing the integrated fluxes of the different
filters. Low-order polynomials have been used to interpolate values,
and uncertainties have been propagated following standard practice.

\begin{figure*}[htp!]
\centering
\includegraphics[width=0.49\linewidth]{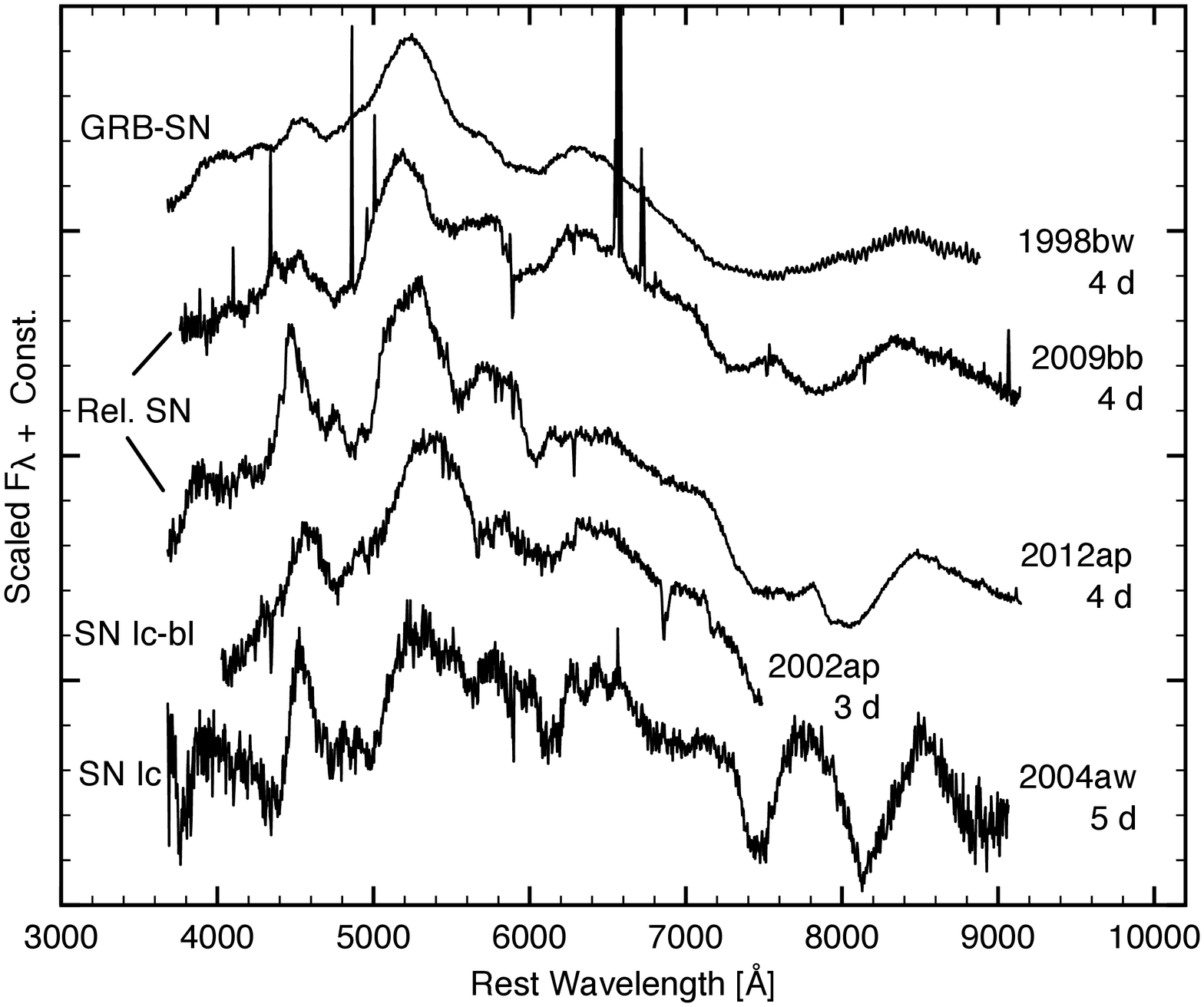}
\includegraphics[width=0.49\linewidth]{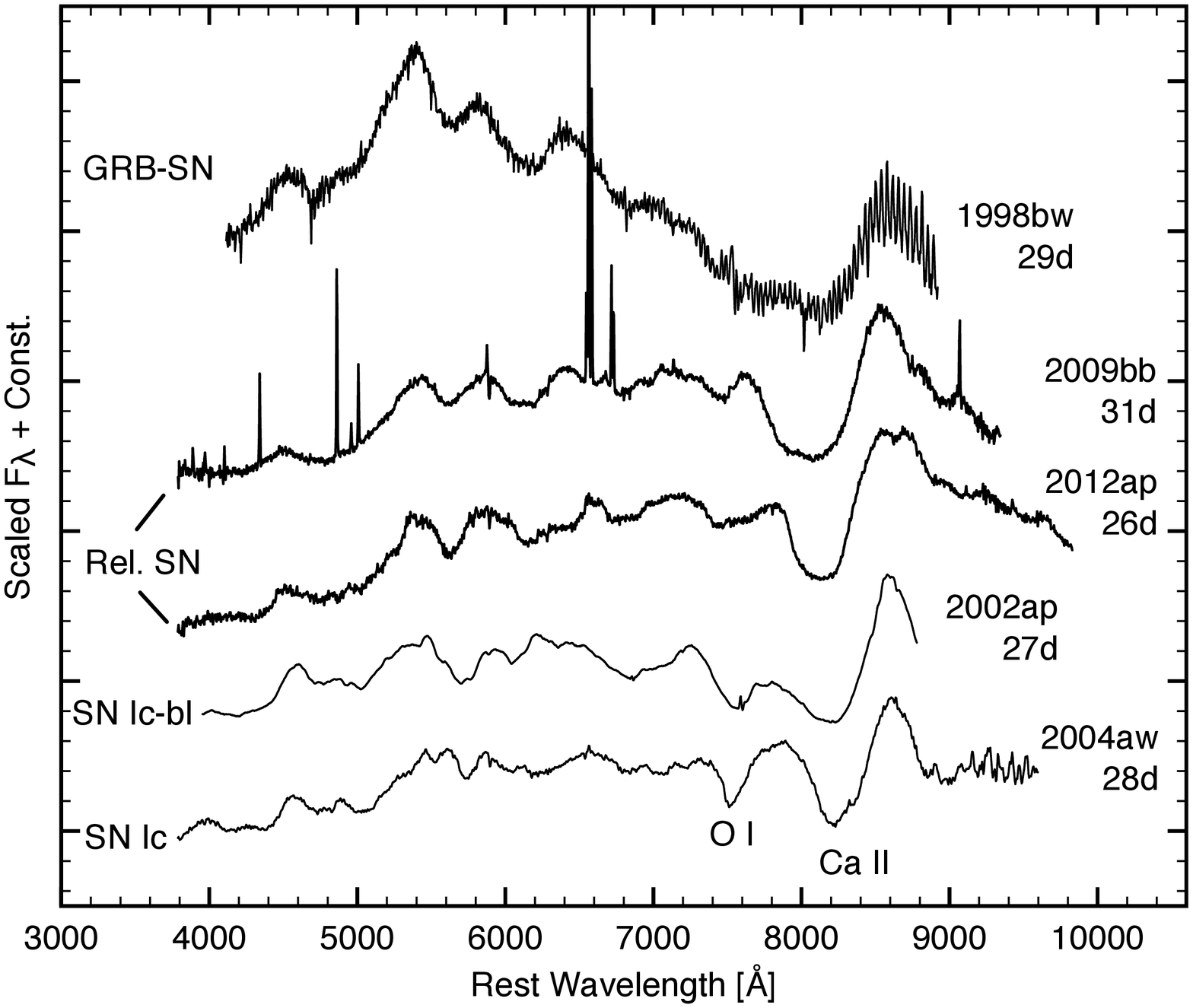}

\caption{Left: Optical spectrum of SN 2012ap around maximum light
  compared to that of various SN\,Ic and Ic-bl. Right: Optical spectrum of 
  SN\,2012ap around one month post-maximum compared to that of various SN\,Ic 
  and Ic-bl. Spectra have been downloaded from the SUSPECT and WISEREP
  databases and were originally published in the following papers:
  SN\,1998bw \citep{Patat01}, SN\,2009bb \citep{Pignata11}, SN\,2002ap
  \citep{Foley03}, and SN\,2004aw \citep{Taubenberger06}.}

\label{fig:speccompare}
\end{figure*}

The quasi-bolometric UV+{\it BVRI} light curve was transformed into a
bolometric light curve assuming $L_{\rm{bol}}^{\rm{quasi}} \approx 0.8\;
L_{\rm{bol}}$, and that $0.2\; L_{\rm{bol}}$ is emitted as unobserved
near-infrared emission. These assumptions follow from observed
properties of SN 2009bb \citep{Pignata11} and are in general agreement
with those observed in other SN\,Ic-bl \citep{Valenti08a}.  The
bolometric light curve is shown in Figure \ref{fig:lbol}. It is worth
noting that outside of SN\,2002ap, SN\,2012ap is among the least luminous
known SN\,Ic-bl.

We modeled the bolometric light curve to derive the ejecta mass
($M_{\rm ej}$), the nickel mass ($M_{\rm Ni}$), and the kinetic energy of the
ejecta ($E_{\rm k}$) following the procedures of \cite{Valenti08a} and
Wheeler, Johnson, \& Clocchiatti (in preparation) that are based on
the formalism outlined by \citet{Arnett82}. We assumed that the
early-time ($\Delta t_{\rm exp} < 30$ d) light curve corresponded with
the photospheric regime, and that the late-time ($\Delta t_{\rm
  exp}>30$ d) light curve corresponded with the nebular regime when
the optical depth of the ejecta decreases and the observed luminosity
is powered by $\gamma$-rays arising from the $^{56}$Co decay,
$\gamma$-rays from electron-positron annihilation, and the kinetic
energy of the positrons (\citealt{Sutherland84,Cappellaro97}). We also
assumed that the rise time was 13 days.  Wheeler, Johnson, \&
Clocchiatti (in preparation) present a detailed outline of all
assumptions and caveats used in the \citet{Arnett82} model that we
have adopted.

Using a conservative estimate of the photospheric velocity of 20,000
\kms\ (see \S \ref{sec:specfeatures}), modeling of our data
yields the following values for the explosion parameters: ${\rm M}_{\rm Ni} =
0.12 \pm 0.02\; {\rm M}_{\odot}$, $E_k = (0.9 \pm 0.3) \times 10^{52}\;\rm
erg$, and $M_{\rm ej}= 2.7 \pm 0.5\; {\rm M}_{\odot}$. The quoted
uncertainties come from a range of optical opacities that were
evaluated from $k_{\rm{opt}}=0.05$--0.1\,$\rm{cm^{2}\,g^{-1}}$.

\subsection{Spectral Evolution}
\label{sec:spectra}

The early optical spectra of SN\,2012ap (Fig.~\ref{fig:specallepochs})
exhibit broad features dominated by \ion{Fe}{2}, \ion{Ca}{2}, and
\ion{O}{1} with velocities starting at about 20,000 \kms\ as measured
from our earliest observation on day $-3.4$. These ions and velocities
are not unlike those observed in many SN\,Ic-bl. In
Figure~\ref{fig:speccompare}, spectra of SN 2012ap observed near the
time of maximum light and around day 30 are shown and compared to
those of various SN\,Ic-bl. Also shown is the peculiar SN\,Ic\,SN
2004aw, which was interpreted as being transitional between SN\,Ic and
SN\,Ic-bl \citep{Taubenberger06}.

Around epochs near maximum light, the spectral features of SN\,2012ap
straddle those observed in the GRB-SN 1998bw and the Type Ic
SN\,2004aw at the extremes. Specifically, the absorptions of SN 2012ap
are not as broad as those observed in SN\,1998bw, nor are they as
narrow, numerous, or weak as those observed in SN\,2004aw. By day 30,
the spectral features show less diversity and the P-Cygni profile of
the \ion{Ca}{2} near-infrared triplet is the most conspicuous feature
in all examples shown. Interestingly, \ion{O}{1} $\lambda$7774 is
stronger in progression from SN\,1998bw to SN\,2004aw, which parallels
the approximate order of decreasing kinetic energy.

\subsection{Spectral Features}
\label{sec:specfeatures}

We utilized the fast and direct P-Cygni summation code \texttt{SYN++}
to assess the atomic makeup of spectral features for SN\,2012ap and
simultaneously extract projected Doppler velocities (see
\citealt{Thomas11} for details of model parameters).  Most ions are
associated with an exponential line optical depth profile starting at
the (assumed sharp) photospheric velocity (PV). Other species are then
``detached'' above the photosphere at high or very high velocities
(HV, VHV) when necessary (see \citealt{Branch06} and
\citealt{Parrent07}). The excitation temperature, \texttt{temp}, has
been fixed to 7000~K, and we utilize the quadratic warping constant
\texttt{a2} (in addition to \texttt{a0} and \texttt{a1}) in order to
reduce the parameter space associated with needing an overly effective
source of line-blanketing blueward of 5000 \AA\ for a given blackbody
continuum level.

\begin{figure*}[htp!]
\centering
\includegraphics[width=0.70\linewidth]{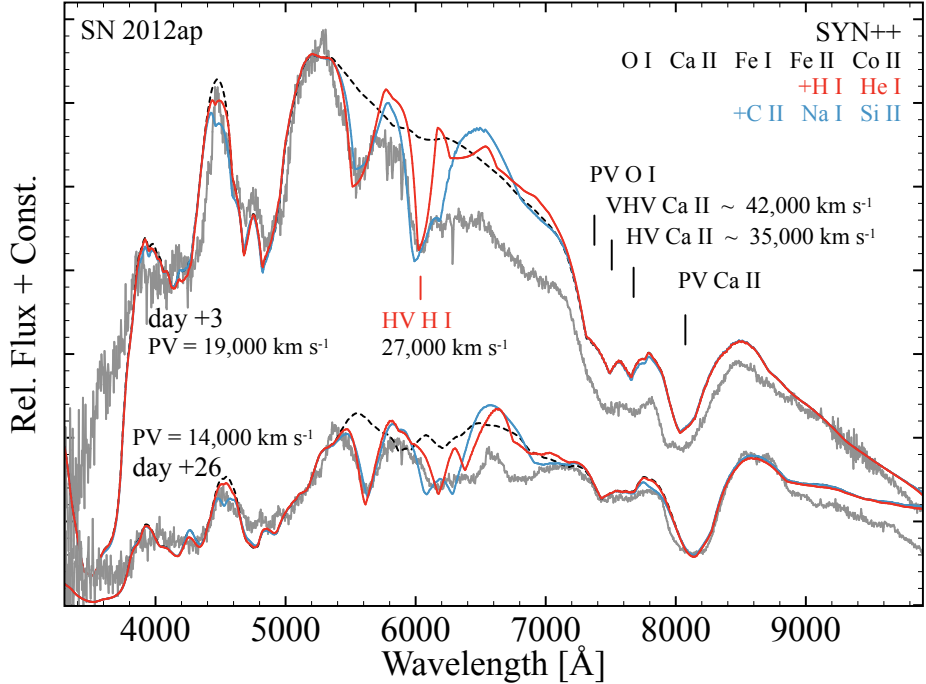}

\caption{\texttt{SYN++} synthetic spectrum comparisons to day $+$3 and
  day $+$26 observations of SN\,2012ap (in grey). Ions considered for
  our fits are shown in the top-right corner. The dashed black line
  shows our ``base model'' of assumed ions, while the red and blue
  lines represent the full fit with \ion{H}{1} and \ion{He}{1}, and
  \ion{C}{2}, \ion{Na}{1}, and \ion{Si}{2}, respectively. All ions
  presented are at the labeled photospheric velocity (PV), unless
  stated otherwise.}

\label{fig:synfit}
\end{figure*}

In Figure~\ref{fig:synfit}, we show a representative \texttt{SYN++}
best fit for a near-maximum and post-maximum light spectrum of
SN\,2012ap. From our analysis, both observed spectra are primarily
consistent with signatures of \ion{Ca}{2} and \ion{Fe}{2} between
19,000 and 14,000 km s$^{-1}$. \ion{Fe}{1} is not explicitly detected;
however, its introduction provides the necessary enhanced
line-blanketing between 4000 and 5000\,\AA\ without conflicting
elsewhere in the fit (e.g., as in the case of
\ion{Co}{1}). Contribution from \ion{Mg}{1} cannot be ruled out. The
small change in measured velocity implies a shallow change in the
photospheric velocity in the 23 days that separate the observations.

Near maximum light, a large absorption trough is
observed at 7000--8500\,\AA. We find fair agreement with multiple
components of \ion{Ca}{2}, including detached components of HV and VHV
\ion{Ca}{2} at 35,000 and 42,000 \kms, respectively. For these
inferred components of Ca-rich material, the fit is convincing so far
as the observed absorption features are not largely consistent with
photospheric \ion{O}{1} and/or \ion{Mg}{2}. Similar to previous
studies of SN\,Ibc spectra, we find a degeneracy between PV \ion{He}{1}
and \ion{Na}{1} for the absorption feature at 5560 \AA\ (see
\citealt{Valenti11}).

\begin{figure}[htp!]
\centering
\includegraphics[width=0.9\linewidth]{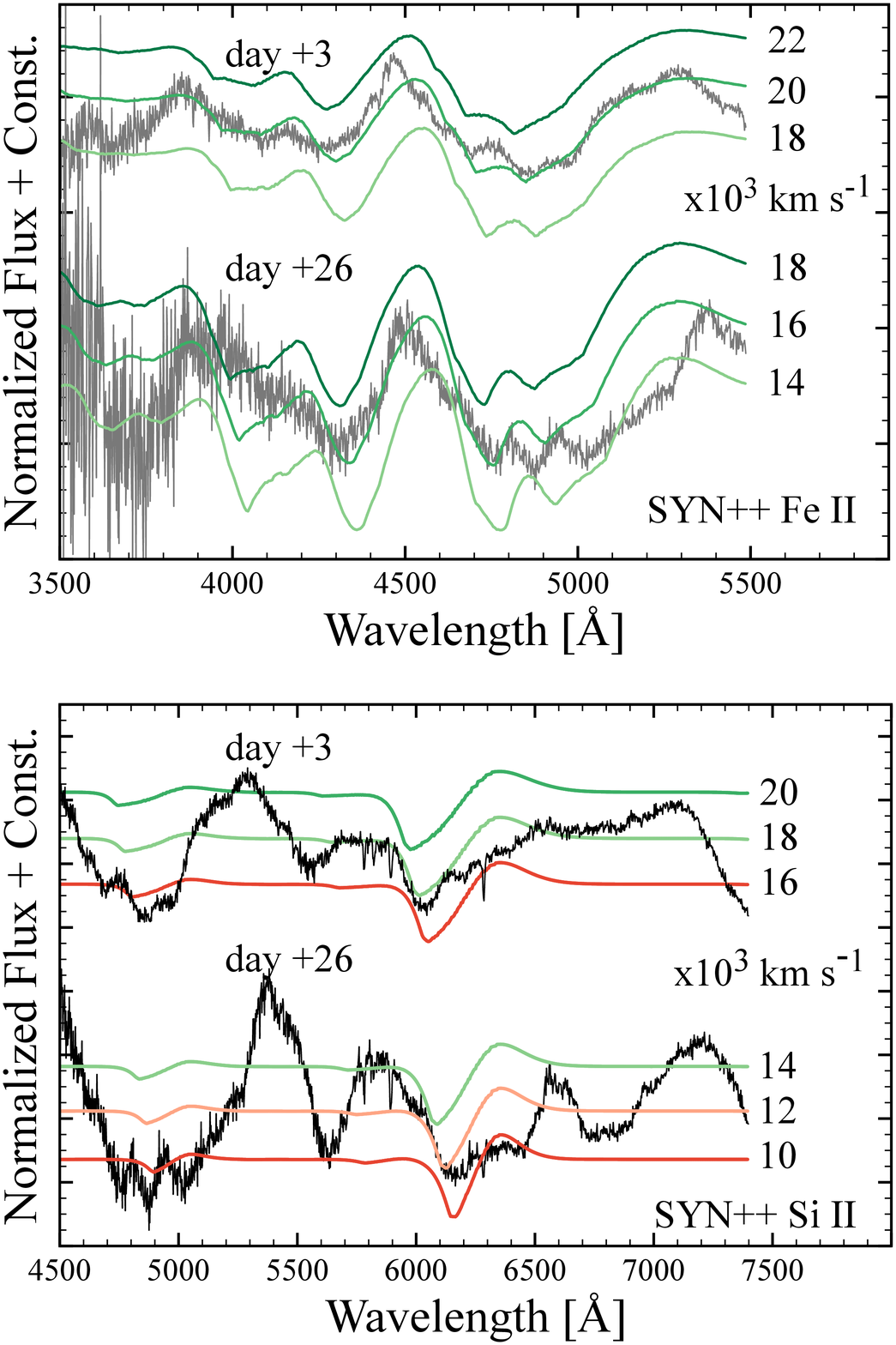}

\caption{Individual ion fits for \ion{Fe}{2} and \ion{Si}{2}. For both
  the day +3 and day +26 spectrum, \ion{Fe}{2} is most consistent with
  projected Doppler velocities of 20,000 \kms\ and 16,000 \kms,
  respectively. The projected Doppler velocities for \ion{Si}{2} at
  these same epochs is $\sim 2000$--4000 \kms\ below those of
  \ion{Fe}{2}. }

\label{fig:fe2test}
\end{figure}

Our best fit for the absorption feature at 6050 \AA\ near maximum
light is detached \ion{H}{1} at 27,000 km s$^{-1}$. As shown in
Figure~\ref{fig:fe2test}, use of PV \ion{Si}{2} for the 6050 \AA\
feature produces absorptions that are consistently too blue throughout
post-maximum epochs. Resolving the fit in favor of \ion{Si}{2} would
require a lower PV, while detaching all other species. This is not a
reasonable physical situation 
(\citealt{Jeffery90,Ketchum08,James10}). Introduction of PV \ion{Si}{2}
improves the overall fit, but it is not a dominant contributor to any
individual feature.

\begin{figure*}[htp!]
\centering
\includegraphics[width=0.75\linewidth]{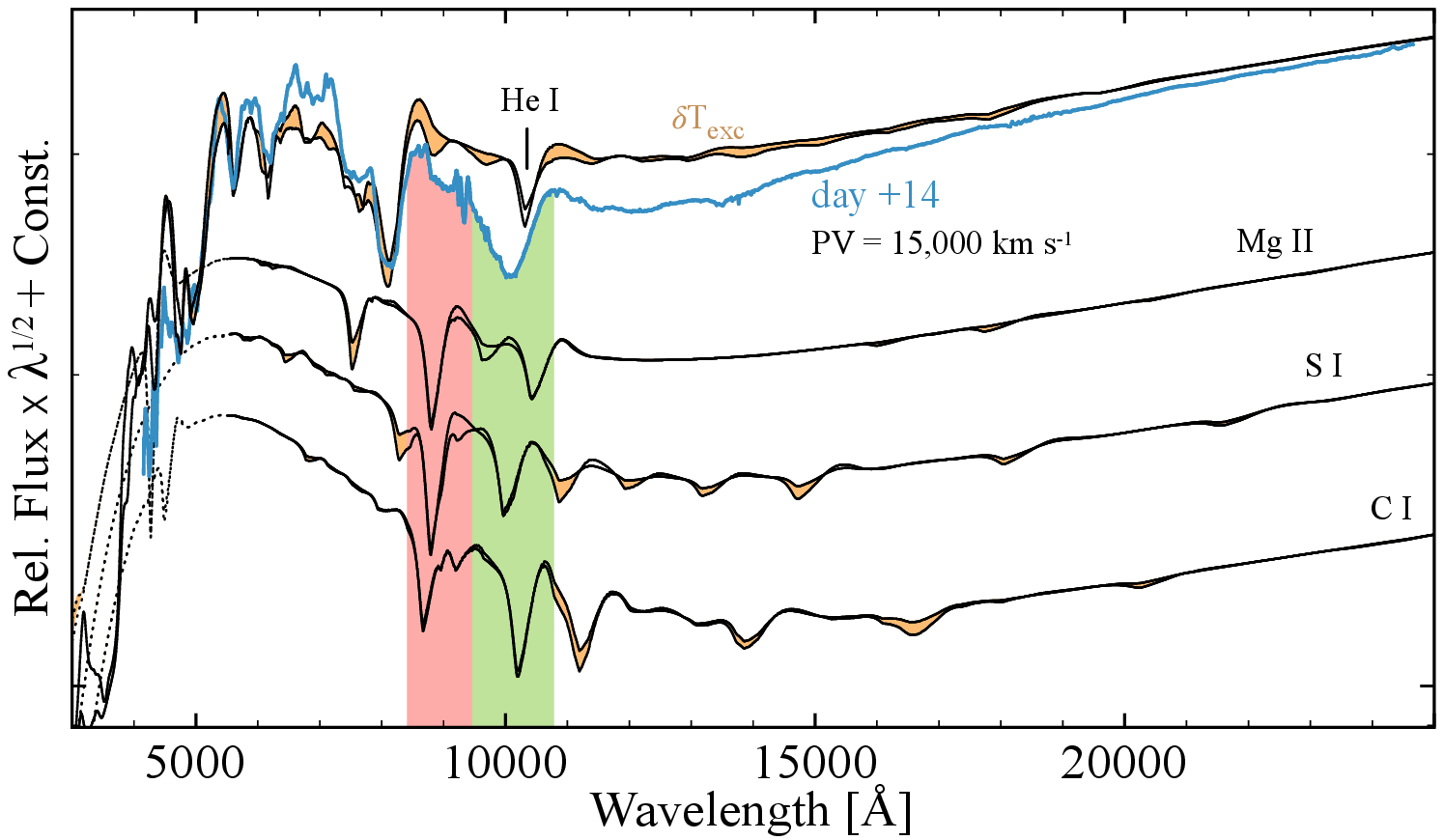}

\caption{Spectral comparisons considering the near-infrared spectrum
  at day $+$14 (blue line). Bands of colors highlight where the
  overlap between the observations and our model is promising (green)
  and in conflict (red). Shown in yellow is the degradation of our
  best \texttt{SYN++} fit when we increase the local thermodynamic
  equilibrium (LTE) excitation
  temperature parameter, $T_{\rm exc}$ (\texttt{temp}), from 7000\,K to
  $10^{4}$\,K. }

\label{fig:syn-nir}
\end{figure*}

In order to test the consistency of our spectroscopic interpretations
between adjacent wavelength regions, in Figure~\ref{fig:syn-nir} we
extend our best optical \texttt{SYN++} fit out to near-infrared
wavelengths using the Magellan/FIRE spectrum obtained on day $+14$. We
find consistency with the inference of \ion{He}{1} at optical
wavelengths and the large 1\,$\mu$m absorption trough, although the
\ion{He}{1} $\lambda$10830 cannot fully account for the breadth of the
1$\mu$m feature. Other ions such as \ion{Mg}{2}, \ion{S}{1}, and
\ion{C}{1} are plausible contributors to the 1$\mu$m absorption, but
each of these species can be immediately ruled out as solely
responsible given respective conflicts at neighboring wavelengths.
Reducing optical depths is not enough to sufficiently hide these
conflicting absorption signatures without invoking non-LTE effects.

An additional test for helium is the presence of the \ion{He}{1}
$\lambda$20587 line \citep{Taubenberger06}. There is evidence of
absorption in the 2\,$\mu$m region of the near-infrared spectrum of
SN\,2012ap, but the S/N is poor and the absorption, if present, would
be at a level where the relative flux ratio $R =
\lambda$20587/$\lambda$10830 $<< 1$. Models favor that if the
\ion{He}{1} $\lambda$10830 line is present, then the \ion{He}{1}
$\lambda$20587 line should also be visible at comparable strength ($R
\approx 1$; \citealt{Mazzali98}).

However, the strength of the \ion{He}{1} lines is expected to increase
with time, as gamma rays penetrate further, since the excited levels
from which \ion{He}{1} lines form are populated almost exclusively by
non-thermal processes \citep{Mazzali98}. Indeed, the strength of the
\ion{He}{1} $\lambda$20587 line has been seen to increase over time
\citep{Marion14}. Since the only near-infrared spectrum of SN\,2012ap
has an epoch of 14 days after maximum, it is possible that the
\ion{He}{1} $\lambda$10830 and $\lambda$20587 lines have not fully
developed yet. The strength of the optical lines of \ion{He}{1} should
be proportional to those in the near-infrared.

We consider the detection of helium-rich ejecta as reasonably viable.
That \ion{He}{1} is unable to solely account for the 1\,$\mu$m
absorption does not rule it out from contributing to this {\it
  compound} feature. It is reasonable to suspect that near-infrared
wavelengths are sampling more layers than accounted for by the
simplified LTE \texttt{SYN++} model
\citep{Wheeler98,Branch02,Dessart11}. Furthermore, non-LTE
calculations with the radiative transfor code PHOENIX (see e.g.,
\citealt{Hauschildt99} and \citealt{Hauschildt14}) have demonstrated
that permitted lines of \ion{Fe}{2} may contribute absorption features
blueward of 1\,$\mu$ \citep{Friesen14}.

\begin{figure*}
\centering
\includegraphics[width=0.65\linewidth]{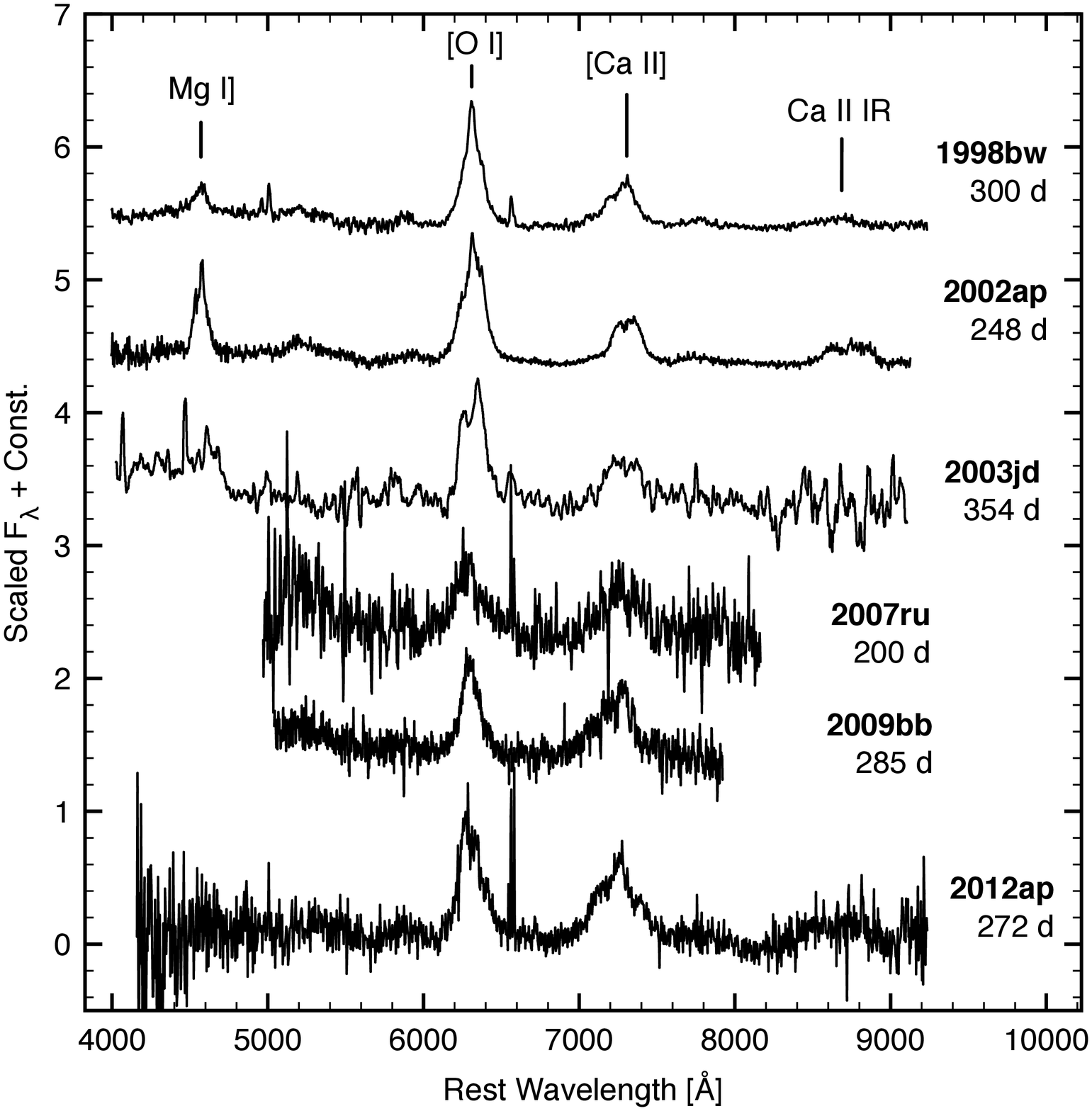}

\caption{Nebular spectrum of SN\,2012ap compared to that of other SN
  Ic-bl. Spectra have been downloaded from the SUSPECT and WISEREP
  databases and were originally published in the following papers:
  SN\,1998bw \citep{Patat01}, SN\,2002ap \citep{Taubenberger09},
  SN\,2003jd \citep{Valenti08}, SN\,2007ru \citep{Sahu09}, and
  SN\,2009bb \citep{Pignata11}. Emission labeled as [\ion{Ca}{2}] has
  potential contributions from [\ion{Fe}{2}] $\lambda$7155 and
  $\lambda$7172, and \ion{Ca}{2} IR has potential contributions from
  [\ion{Fe}{2}] $\lambda$8617 and [\ion{C}{1}] $\lambda$8727.}

\label{fig:nebspeccompare}
\end{figure*}

\subsection{Analysis of Nebular Spectra}
\label{sec:nebspec}

Optical spectra of SN\,2012ap during the nebular phase ($t > 200$\,d;
Figure \ref{fig:latetimespec}) exhibit conspicuous emissions
associated with [\ion{O}{1}] \dlambda 6300, 6364, [\ion{Ca}{2}]
\dlambda 7291, 7324, and the \ion{Ca}{2} triplet around 8600\,\AA. These
emissions originating from inner metal-rich ejecta heated by
radioactive $^{56}$Co are typical of SN\,Ibc at late stages
\citep{Filippenko90,Mazzali01,Matheson01,Taubenberger09,Modjaz14}.
Other emission lines that are also often observed at these epochs are
weakly detected, including \ion{Mg}{1}] $\lambda$4571, \ion{Na}{1}~D,
and additional emission around 5200 \AA\ associated with blends of
[\ion{Fe}{2}].

\begin{figure}
\centering
\includegraphics[width=0.8\linewidth]{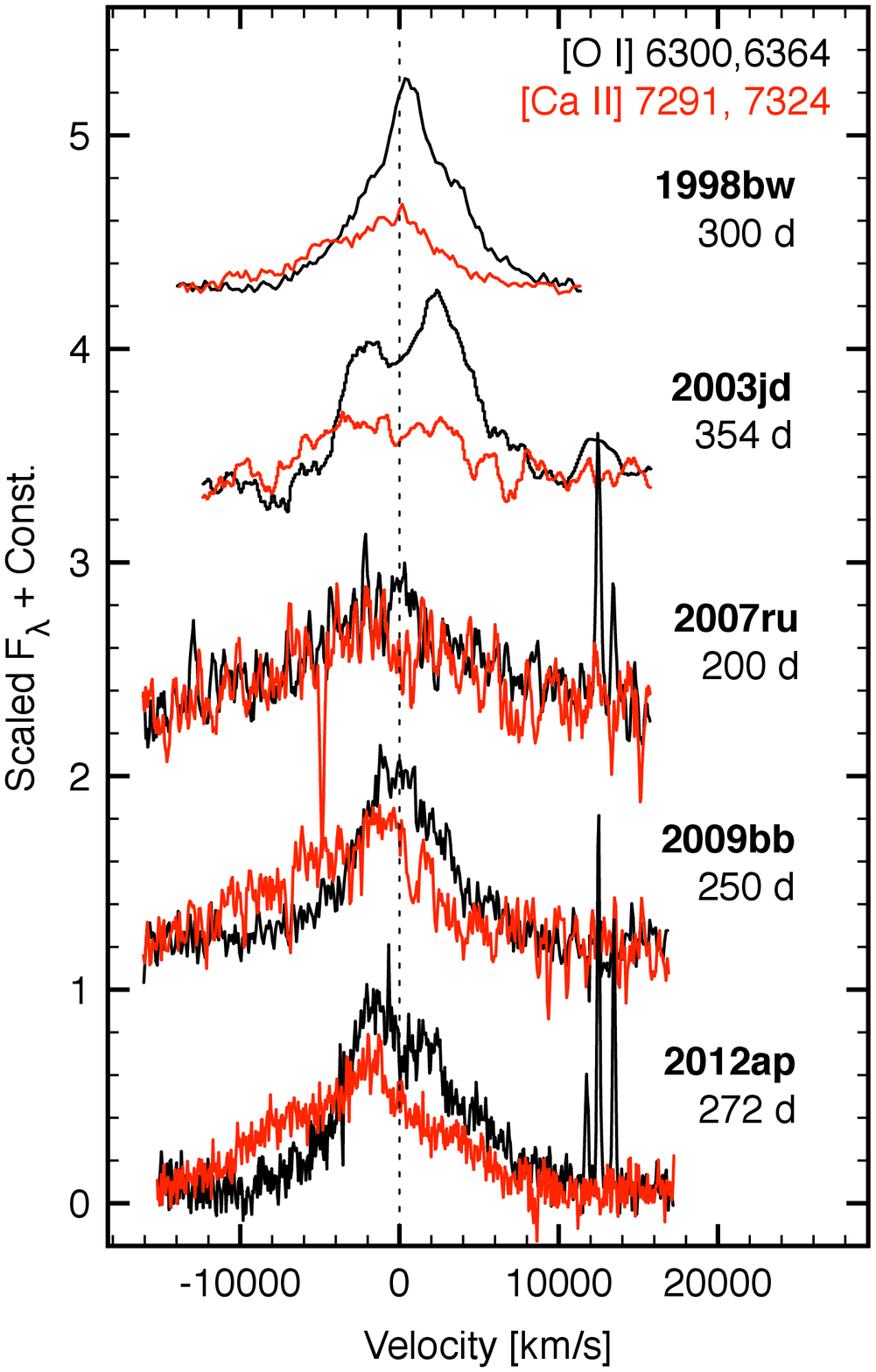}

\caption{Emission-line profiles of [\ion{O}{1}] \dlambda 6300, 6364
  and [\ion{Ca}{2}] \dlambda 7291, 7324 observed in SN\,2012ap and
  other broad-lined Type Ic supernovae from Figure
  \ref{fig:nebspeccompare}. Velocities for [\ion{O}{1}] are with
  respect to 6300~\AA, and velocities for [\ion{Ca}{2}] are with
  respect to 7306~\AA. Flux values for [\ion{O}{1}] are normalized to
  one and the relative flux between [\ion{O}{1}]/[\ion{Ca}{2}] has
  been maintained. [\ion{Ca}{2}] profiles are contaminated to varying
  degrees by the [\ion{Fe}{1}] $\lambda$7155 and $\lambda$7172 lines.}

\label{fig:OICaIIvelocity}
\end{figure}

Figure \ref{fig:nebspeccompare} shows a representative late-time
spectrum of SN\,2012ap compared to spectra of other SN\,Ic-bl. The
\ion{Mg}{1}] $\lambda$4571 line is particularly weak in the day 272
spectrum, although our observations do suffer from reduced sensitivity
in this wavelength region.  The relative line flux ratio of
[\ion{Ca}{2}]/[\ion{O}{1}] $\approx 1.2$ is large in SN 2012ap
compared to those observed in SN\,1998bw and SN\,2002ap where it is
$\approx 0.5$. SN\,2009bb and SN\,2007ru also show
[\ion{Ca}{2}]/[\ion{O}{1}] $> 1$. This line flux ratio has been
suggested to be a useful indicator of progenitor core mass, with
larger [\ion{Ca}{2}]/[\ion{O}{1}] ratios indicative of a less massive
helium core at the time of explosion \citep{Fransson89}.

Emission-line velocity plots comparing the lines in SN\,2012ap to
those of other SN\,Ic-bl are shown in
Figure~\ref{fig:OICaIIvelocity}. The observed [\ion{O}{1}] and
[\ion{Ca}{2}] emissions exhibit line velocities of $\sim 5500$ \kms\,
that are not unlike those observed in other SN\,Ic-bl
\citep{Maurer10-Velocity}. The [\ion{Ca}{2}] \dlambda 7291, 7324
emission is noticeably broad, and its blueshifted velocities appear to
be greater than those of oxygen. However, emission blueward of 7306
\AA\ (which we assume to be the center of the distribution) has likely
contribution from the [\ion{Fe}{2}] $\lambda$7155 and $\lambda$7172
lines. The single peak of the entire [\ion{Ca}{2}] distribution is
blueshifted by 1700 \kms\ and has a sharp dropoff on the redshifted
side.

The [\ion{O}{1}] emission appears to be double-peaked. The blueshifted
and strongest peak in the distribution is centered near $-1700$ \kms,
which is the same velocity as the peak observed in the [\ion{Ca}{2}]
distribution. The minimum between the two peaks is near 6300 \AA\
($\approx 0\;$\kms). Between days 218 and 272, emission redward of
6300 \AA\ increases in strength. This evolution, illustrated in
Figure~\ref{fig:latetimespecOI}, is discussed further in \S
\ref{sec:kin}.

We modeled the nebular spectrum using our SN nebular spectrum code,
assuming that the late-time emission is tied to the deposition of
gamma-rays and positrons from $^{56}$Co decay. Given an ejected mass,
a characteristic boundary velocity (which corresponds to the half width 
at half-maximum intensity of the emission lines), and a composition,
the code computes gamma-ray deposition, follows the diffusions of the
gamma-rays and the positrons with a Monte Carlo scheme, and computes
the heating of the gas. The state of the gas is then computed in non-LTE,
balancing heating and cooling via line emission. The code has been
used for a number of SN\,Ibc
(e.g., \citealt{Mazzali01,Mazzali07,Mazzali10}) and is the latest
version described in some detail by \citet{Mazzali11}.

\begin{figure}
\centering
\includegraphics[width=\linewidth]{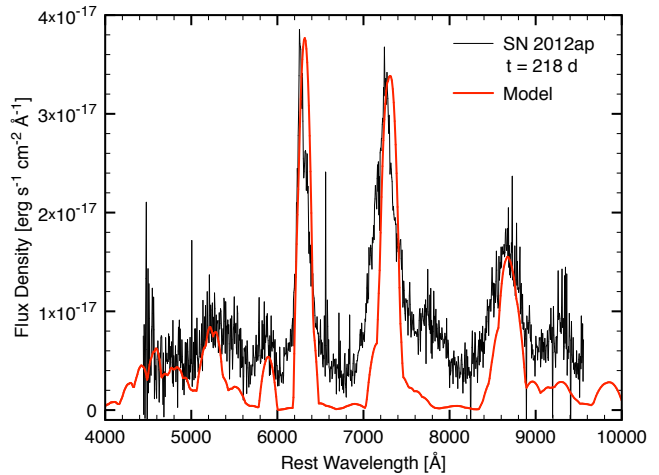}

\caption{Late-time optical spectrum of SN\,2012ap obtained 2012
  September 23 (t = 218 d; black) and our model fit (red).}

\label{fig:nebspecmodel}
\end{figure}

The synthetic spectrum produced by our model for the day 218 optical
spectrum is shown in Figure~\ref{fig:nebspecmodel}. For material
inside 5500 \kms\ we derive a nickel mass of ${\rm M}_{\rm Ni} = 0.20
\pm 0.05\; {\rm M}_{\odot}$. This estimate of ${\rm M}_{\rm Ni}$ is
larger than that derived from modeling of the bolometric light curve
in \S \ref{sec:lc} ($0.12 \pm 0.02\;{\rm M}_{\odot}$), but not grossly
inconsistent. Additionally, we estimate the oxygen mass to be ${\rm
  M}_{\rm oxygen} \approx 0.5\;{\rm M}_{\odot}$ and the total ejecta
mass to be ${\rm M}_{\rm ej} \approx 0.8\;{\rm M}_{\odot}$. The value
of $M_{\rm ej}$ calculated by our models of the nebular spectra is
less than that derived in \S \ref{sec:lc} ($\sim 2.7\; {\rm
  M}_{\odot}$). Some of the discrepancy is because this value is only
for mass inside 5500 \kms. Also, as we show in \S \ref{sec:kin}, there
is a possibility of internal absorption.  If the emission lines are
not optically thin, then the masses derived from them will be
underestimated.

\section{Discussion}
\label{sec:discussion}

Our ultraviolet, optical, and near-infrared observations of SN\,2012ap
show it to be a member of the energetic SN\,Ic-bl class with explosion
properties that fall in between normal SN\,Ibc events and
SN-GRBs. SN\,2012ap follows SN\,2009bb as the second example of a 
SN\,Ic-bl associated with ejecta moving at relativistic velocities ($v \ga
0.6\;\rm c$) but not associated with a gamma-ray burst detection. In
the case of SN 2009bb, its relativistic ejecta continued to be in
nearly free expansion for $\sim 1$\,yr, which is unlike
GRBs. SN\,2012ap's relativistic ejecta, however, did slow down on
timescales similar to those of GRBs \citep{Chakraborti14}.

A very small percentage of supernova explosions ($<1$\%) harbor a
central engine capable of powering an ultra-relativistic jet detected
as a GRB \citep{Soderberg06}. Thus, it is of interest to compare and
contrast the properties of SN\,2009bb and SN\,2012ap to those of other
SN\,Ibc and GRB-SN. Below we discuss some of the more interesting
observational properties of these relativistic supernovae without a
GRB detection uncovered in our analysis, and discuss their relevance
to potential progenitor systems and explosion mechanisms.

\subsection{Hydrogen and Helium in SN\,Ic-bl}
\label{sec:kin}

Our multi-epoch \texttt{SYN++} fits of the optical and near-infrared
spectra of SN\,2012ap strongly favor the identification of helium in
the ejecta. Detection begins from the first observation on day $-3.4$
and photospheric velocities of 20,000 \kms\ are measured.  In addition,
hydrogen is likely to be present and detached above the photosphere traveling
at $\sim 27,000$ \kms. Helium was also detected in pre-maximum light
spectra of SN\,2009bb \citep{Pignata11}, though in that case it was
not as conspicuous, and hydrogen was not reported.

It is not unusual for helium to be detected in the spectra of SN\,Ic
and SN\,Ic-bl. Observations of sufficient quality and temporal
coverage in combination with spectral modeling have shown that
absorption features in a variety of SN\,Ibc spectra can sometimes be
best understood as arising from helium
\citep{Filippenko88,Filippenko92,Filippenko95,Clocchiatti96-94I,Patat01,Mazzali02,Elmhamdi06,Branch06,Bufano12}.
In some of these instances, hydrogen traveling at velocities above the
photosphere has also been identified as a possible constituent of the
ejecta.

However, the extent of H and He in SN\,Ibc, and especially SN\,Ic-bl,
remains debated \citep{Matheson01,Branch02,Branch06,Hachinger12,
Milisavljevic13,Modjaz14}.  It is particularly difficult to derive masses of
He because of the high excitation potentials of He that exceed the
energy of thermal photons and electrons, and require detailed non-LTE
spectral modeling.  The presence or lack of \ion{He}{1} lines in SN\,Ibc 
may indicate a genuine helium deficiency, or, alternatively, the
result of inadequate excitation
\citep{Harkness87,Dessart11,Li12}. Only a small amount of helium ($\la
0.1\; {\rm M}_{\odot}$) needs to be present to be observable
\citep{Hachinger12}, but asymmetric mixing or weak mixing may prevent
sizable amounts of He from being excited by Co
\citep{Dessart12}. With time, SN expansion will thin the ejecta and
should eventually expose any He-rich material to radioactive decay
\citep{Swartz93}.

Determining the amount of H and He in SN ejecta is an important
constraint on the evolutionary state of the progenitor star at the
time of explosion.  There is ample reason to believe that sizable
quantities of these elements should remain, as no completely
hydrogen-free low-mass helium stars are known \citep{Eldridge13}, and
stellar models have difficulties explaining strong loss of He
\citep{Georgy09,Yoon10}. Recent stellar evolution calculations predict
that the mass of He can reach upwards of $0.5\;{\rm M}_{\odot}$ in SN\,Ibc
progenitors.

\begin{figure}
\centering
\includegraphics[width=\linewidth]{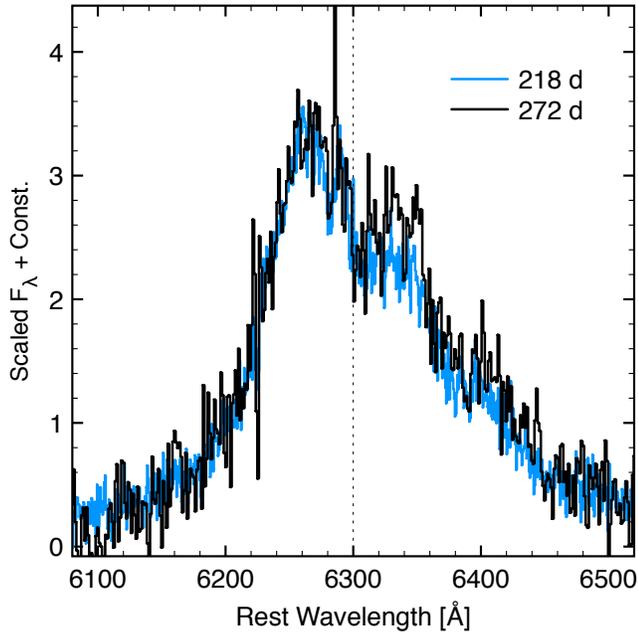}

\caption{Late-time emission of [\ion{O}{1}] \dlambda 6300, 6364 in
  SN\,2012ap betwen day 218 and 272. Emission redward of 6300~\AA\
  increases with time.}

\label{fig:latetimespecOI}
\end{figure}

\subsection{Asymmetric [\ion{O}{1}] Emission}

Conspicuous asymmetry in the [\ion{O}{1}] \dlambda 6300, 6364 profile
of SN\,2012ap is suggestive of two peaks
(Fig.~\ref{fig:latetimespecOI}). Examinations of late-time spectra of
many SN\,Ibc ($\ga 40$) have demonstrated that double-peaked line
profiles in [\ion{O}{1}] emission seem to be a relatively common
phenomenon occurring in 30--40\% of objects
\citep{Maeda08,Modjaz08,Taubenberger09,Milisavljevic10}. Such
double-peaked emission-line profiles deviate from the single-peaked
profile expected from a spherically symmetric source, and this has been
interpreted as evidence for aspheric debris having a toroidal or
disk-like geometry \citep{Maeda02}.  \citet{Mazzali05} observed a
double-peaked [\ion{O}{1}] profile in the SN\,Ic-bl 2003jd and
interpreted it as emission originating from a torus of O-rich debris
perpendicular to a high-velocity jet in a GRB model. In this
framework, nondetections of GRBs from SN\,Ic-bl are the consequence of
viewing jet-driven explosions along the equatorial expansion plane
perpendicular to the jet axis.

The double-peaked [\ion{O}{1}] profile observed in SN\,2012ap could be
produced by a toroidal ejecta geometry similar to the Mazzali et al.\
model. If so, the distribution of O-rich material in SN\,2012ap was
not as aspheric as in SN\,2003jd because the peaks observed in
SN\,2012ap are neither as pronounced nor as widely separated (see
Fig.~\ref{fig:OICaIIvelocity}). Notably, SN\,2009bb does not exhibit
considerable asymmetry in its [\ion{O}{1}] distribution, although
\citet{Pignata11} report that the signature of aspheric explosion
dynamics might have been seen in earlier nebular data.

Alternatively, the observed [\ion{O}{1}] profile of SN\,2012ap could
be the result of absorption in the SN interior. The evolution in the
relative strength of the two peaks of [\ion{O}{1}] emission is
consistent with this notion. The two peaks are divided by a steep 
dropoff in emission near zero velocity, and the strength of emission
redward of zero velocity increases over time
(Fig.~\ref{fig:latetimespecOI}).  The fact that [\ion{Ca}{2}] is
asymmetric and preferentially blueshifted with a distribution similar
to that observed in [\ion{O}{1}] is also consistent with a significant
portion of interior ejecta being opaque
(Fig.~\ref{fig:OICaIIvelocity}).

Possible contributors to this opacity are high densities or dust in
the ejecta.  Optically thick inner ejecta could prevent light from the
rear side of the SN from penetrating, creating a flux deficit in the
redshifted part of emission lines
\citep{Taubenberger09,Milisavljevic10}, as observed in the late-time
emission profiles of SN\,2012ap (Figure
\ref{fig:OICaIIvelocity}). Internal absorption associated with high
densities is generally not anticipated at these late epochs ($t \ga
300$ d) because SN expansion should lead to low densities with
associated low optical opacities \citep{Maeda08}. However, deviations
from spherical symmetry and clumping make it possible for portions of
the ejecta to remain optically thick
\citep{Spyromilio90,Li92,Chugai94}. 

Dust formation is not directly supported by our observations. There is
no change in the blueshifted position of the peaks toward smaller
wavelengths \citep{Lucy89}, and no late-time infrared observations are
available to detect potential flux excess at wavelengths that would be
associated with heated dust emission (see, e.g.,
\citealt{Dwek83,Kozasa91,Gerardy02,Fox09}).

\subsection{Host Environment and Progenitor System}

Several properties of SN\,2012ap and SN\,2009bb distinguish them from
most SN\,Ic-bl and may have implications for the progenitor systems of
relativistic SN lacking a GRB detection.  One such property is their
host environment metallicities. The oxygen abundance of the explosion
site of SN\,2012ap using the N2 scale of \citet{PP04} is $\log({\rm
  O/H})+12=8.79$ (M14), which is slightly above solar (Z$_{\odot}
\equiv \log({\rm O/H})_{\odot} + 12 = 8.69$; \citealt{Asplund05}). The
host metallicity of SN\,2009bb was estimated to be $1.7-3.5\;{\rm
  Z}_{\odot}$ \citep{Levesque10}. These metallicities are at the high
end of environments of the majority of SN\,Ic-bl discovered by
targeted surveys, and substantially higher than those of SN discovered
in untargeted surveys where the mean metallicity is $\sim {\rm
  Z}_{\odot}/2$
\citep{Anderson10,Modjaz11,Kelly12,Sanders12}. SN\,2009bb and
SN\,2012ap were both discovered by targeted surveys.

It is worth noting that both SN were located in the outer regions of
their host galaxies along the edges of spiral arms. This runs counter
to recent studies that find that the majority of SN\,Ic are much more
likely to occur in the brightest regions of their host galaxies
\citep{Kelly08,Kelly14}. Spiral galaxies typically show decreasing
metallicity gradients with respect to radial distance from the core
\citep{Zaritsky10}, but this trend is known to have important
exceptions \citep{Zurita12}.

Wolf-Rayet (WR) stars have long been a suspected progenitor system of
SN\,Ibc \citep[][and references therein]{Gaskell86,Gal-Yam14}.  
However, considerable evidence, including
luminosity limits derived from nondetections of progenitor stars in
high-resolution pre-explosion images \citep{vanDyk03,Smartt09}, have
ruled out a large population of WR stars, and it is unlikely that WR
stars are the dominant progenitor channel. Instead, relatively low-mass 
helium stars in binary systems may represent a large fraction of SN\,Ibc
\citep{Eldridge08,Smith11,Eldridge13}.

In the case of SN\,2012ap, there is circumstantial evidence in support
of a WR star association. M14 report that the unusually strong DIBs in
the optical spectra that exhibit changes in EW over short ($\la 30$
days) timescales do so in a manner consistent with the SN interacting
with a nearby source of the DIBs. The only other reported cases of
time-varying DIB features have been in families of WR and luminous
blue variable (LBV) stars \citep{LeBertre93,Heydari93}.

Interestingly, SN\,2009bb was also found to be associated with strong
DIB absorptions in its moderately extinguished optical spectra, as was
SN\,2008D, which exhibited broad-lined features similar to SN\,2012ap
before maximum light but then evolved into a SN\,Ib. Although chance
alignments between DIB carrier-rich molecular clouds and these SN are
possible, it may be that the SN progenitor systems are related to the
sources of the DIBs, and that the source could be related to mass loss
from the progenitor star. LBV and WR stars exhibit varying degrees of
asymmetric mass loss \citep{Nota95}; thus, an observer's line of sight
with respect to a circumstellar disk could be an important factor in
explaining why strong DIB detections are rare.

\section{Conclusions}
\label{sec:conclusions}

We have presented ultraviolet, optical, and near-infrared observations
of the broad-lined Type Ic SN\,2012ap, which is the second known
example of a SN from a stripped-envelope progenitor capable of
accelerating a non-negligible portion of its ejecta to relativistic
velocities ($v \ga 0.6\;\rm c$) but not associated with a GRB
detection. The only known counterpart is SN\,2009bb.

Notable properties of these two relativistic SN\,Ic-bl may reflect key
aspects of their progenitor systems and explosion dynamics. In
particular, both SN\,2012ap and SN\,2009bb share the following
characteristics:

\begin{my_itemize}

\item strong radio emission consistent with ejecta being accelerated
  to relativistic velocities but no GRB counterpart,

\item weak X-ray emission at late times ($t > 10$\,d),

\item evidence for helium in early-time optical spectra with
  photospheric velocities of $\ga$$\;$20,000 \kms,

\item relatively large emission-line flux ratio of
  [\ion{Ca}{2}]/[\ion{O}{1}] $ > 1$ in nebular spectra,

\item high levels of internal host extinction ($E(B-V) > 0.4$ mag), and

\item environments of solar to super-solar metallicity, and locations
  along the outer spiral arms of their host galaxies.

\end{my_itemize}

\citet{Margutti14} propose that SN\,2009bb and SN\,2012ap represent
the weakest of central-engine-driven explosions, and conclude that
these events lack an associated GRB detection because engine activity
stops before the jet is able to pierce through the stellar
envelope. Though ``choked,'' the jet is still able to accelerate a
small fraction of ejecta to relativistic velocities. Engines of short
durations or progenitors of large stellar envelopes may inhibit the
jet from completely piercing the surface of the star.

Indeed, SN\,2009bb and SN\,2012ap may be among the weakest explosions
for which we are able to detect the presence of jet activity. The
continuum of explosions extending from GRB-SN to more ordinary SN\,Ibc
suggests that a wider variety of jet activity may potentially be
occurring at energies that are observationally ``hidden.''  Detection
at weaker scales is challenging since such explosions do not produce
electromagnetic signatures as easily recognizable as GRBs. SN of this
variety can be dynamically indistinguishable from ordinary
core-collapse SN \citep{Tan01,Matzner03,Lazzati12}, and/or their
high-energy emissions may be below the threshold of the current
generation of gamma-ray instruments \citep{Pignata11}.

Some hints in support of jets at smaller energy scales come from
supernova remnants.  Cassiopeia A, known to be the result of an
asymmetric Type IIb explosion \citep{Krause08,Rest11}, exhibits
exceptionally high velocity Si- and S-rich material in a
jet/counter-jet arrangement
\citep{Fesen01,Hines04,Hwang04,Fesen06}. The known extent of this jet
region contains fragmented knots of debris traveling $\sim$$\;$15,000
\kms, which is three times the velocity of the bulk of the O- and
S-rich main shell. Though the large opening half-angle of this
high-velocity ejecta is inconsistent with a highly collimated flow
\citep{Milisavljevic+Fesen13}, some jet-like mechanism carved a path
allowing interior material from the Si-S-Ar-Ca region near the core
out past the mantle and H- and He-rich photosphere. This process,
potentially related to a protoneutron star wind that follows the
supernova outburst \citep{Burrows05-Rotate}, would be observationally
indistinguishable from non-jet explosion models at extragalactic
distances.

Rotation is believed to be a key variable driving the outcome of these
explosions. If the jet is associated with the protoneutron star or
magnetar wind that follows the SN, rapid rotation will naturally
amplify magnetic fields and make magnetohydrodynamic power influential
\citep{Akiyama03}. GRBs may only come from the most rapidly rotating
and most massive stars \citep{Woosley06,Burrows07}. Progenitor
composition and structure is another important consideration in SN
explosions \citep{Arnett11,Ugliano12}. He and/or H layers that can
vary in thickness along different viewing angles \citep{Maund09}, have
the potential to quench relativistic jets (e.g., SN\,2008D;
\citealt{Mazzali08}), and can lead to expansion asymmetries that can
be potentially detected by spectropolarimetry
\citep{Maund07,Wang08,Tanaka08,Tanaka12}. The inferred presence of
helium in the optical spectra of SN\,2012ap and SN\,2009bb is
consistent with the quenched jet scenario.

With only two events identified so far, it is possible that the
exceptional properties of SN\,2012ap and 2009bb discussed here are
biased by transient surveys targeting metal-rich systems. Additional
examples of relativistic SN\,Ic-bl are needed to test whether the
properties maintain and to further understand these events in the
entire context of SN\,Ibc. Given their complex origins --- e.g.,
line-of-sight influences of potentially asymmetric explosions inside
progenitors of varying structure and compositions, varying
core-rotation speeds, and asymmetric circumburst mediums of differing
metallicities --- real progress requires an in-depth and
multi-wavelength (radio through gamma-rays) approach studying a large
sample ($N > 30$) of local SN\,Ic-bl.

\acknowledgements

Many of the observations reported in this paper were obtained with the
Southern African Large Telescope. Additional data presented herein
were obtained at the W.~M. Keck Observatory, which is operated as a
scientific partnership among the California Institute of Technology,
the University of California, and NASA; the observatory was made
possible by the generous financial support of the W.~M.\ Keck
Foundation. Some observations also came from the MMT Observatory, a
joint facility of the Smithsonian Institution and the University of
Arizona, as well as the 6.5 m Magellan Telescopes located at Las
Campanas Observatory, Chile. Support was provided by the David and
Lucile Packard Foundation Fellowship for Science and Engineering
awarded to A.M.S. J.M.S.\ is supported by an NSF Astronomy and
Astrophysics Postdoctoral Fellowship under award AST-1302771. T.E.P.\
thanks the National Research Foundation of South Africa. R.P.K. and
J.C.W. are grateful for NSF grants AST-1211196 and AST-1109801,
respectively. A.V.F.\ and S.B.C.\ acknowledge generous support from
Gary and Cynthia Bengier, the Richard and Rhoda Goldman Fund, the
Christopher R. Redlich Fund, the TABASGO Foundation, and NSF grant
AST-1211916. K.M.\ acknowledges financial support by Grant-in-Aid for
Scientific Research for Young Scientists (23740141, 26800100). The
work by K.M.\ is partly supported by World Premier International
Research Center Initiative (WPI Initiative), MEXT, Japan. J.V.\ is
supported by Hungarian OTKA Grant NN-107637. M.D.S. and E.Y.H.\
gratefully acknowledge generous support provided by the Danish Agency
for Science and Technology and Innovation realized through a Sapere
Aude Level 2 grant. E.Y.H.\ also aknowledges support from NSF grant
AST-1008343. D.\ Sahu and G.\ Pignata kindly provided archival spectra
of SN 2007ru and SN 2009bb, respectively. This paper made extensive
use of the SUSPECT database (\texttt{www.nhn.ou.edu/$\sim$suspect/})
and the Weizmann interactive supernova data repository
(\texttt{www.weizmann.ac.il/astrophysics/wiserep}). This work was
supported in part by NSF Grant No.\ PHYS-1066293 and the hospitality
of the Aspen Center for Physics.

\clearpage
\newpage


\begin{deluxetable*}{lcccc}
\footnotesize
\centering
\tablecaption{KAIT photometry (mag) of SN 2012ap}
\tablecolumns{5}
\tablewidth{0pt}
\tablehead{\colhead{MJD}    &
           \colhead{$B$}   &
           \colhead{$V$}   &
           \colhead{$R$}    &
           \colhead{$I$}      
}
\startdata
55967.25	& \nodata	&17.46	0.08	&17.23    0.10 &\nodata \\
55973.15	& 17.77	0.04	& 16.74	0.03	&16.44	0.04	&16.00	0.04\\
55974.19	& 17.65	0.03	& 16.77	0.02	&16.36	0.02	&16.00	0.02\\
55978.15	& 17.88	0.04	& 16.76	0.01	&16.29	0.02	&15.85	0.02\\
55980.13	& 18.15	0.06	& 16.78	0.03	&16.35	0.02	&15.76	0.02\\
55980.17	& \nodata	& \nodata	&16.36	0.04 & \nodata \\
55981.15	& 18.32	0.08	& 16.90	0.03	&16.35	0.03	&15.75	0.04\\
55982.14	& 18.32	0.07	& 16.94	0.04	&16.43	0.04	&15.77	0.06\\
55983.14	& 18.39	0.07	& 16.98	0.03	&16.44	0.02	&15.78	0.02\\
55990.14	& 18.97	0.18	& 17.56	0.06	&16.81	0.04	&16.10	0.02\\
55990.18	& \nodata	& \nodata	&16.81	0.08 & \nodata\\
55991.14	& 18.91	0.10	& 17.65	0.05	&16.94	0.04	&16.18	0.05\\
55992.16	& 19.12	0.28	& 17.63	0.07	&16.96	0.04	&16.27	0.04\\
55993.14	& 18.95	0.22	& 17.80	0.07	&17.02	0.05	&16.21	0.05\\
55994.15	& 19.16	0.29	& 17.89	0.13	&17.14	0.08	&16.29	0.05\\
55995.16	& 19.55	0.18	& 17.93	0.06	&17.11	0.03	&16.31	0.03\\
55996.15	& 19.51	0.13	& 17.92	0.05	&17.20	0.03	&16.41	0.03\\
55998.19	& 19.40	0.19	& 18.06	0.14	&\nodata  	& 16.51	0.06\\
\enddata

\label{tab:kaitphot}
\end{deluxetable*}

\begin{deluxetable*}{lcccc}
\footnotesize
\centering
\tablecaption{Swift-UVOT photometry (mag) of SN 2012ap}
\tablecolumns{5}
\tablewidth{0pt}
\tablehead{\colhead{MJD}    &
           \colhead{$w1$}         &
           \colhead{$u$}           &
           \colhead{$b$}           &
           \colhead{$v$}}
\startdata
55969.26	& 19.30	0.23	& 18.30	0.14	& 17.81	0.08	& 17.05	0.08\\
55971.24	& 18.88	0.12	& 17.91	0.09	& 17.65	0.06	& 16.90	0.07\\
55973.41	& 19.06	0.13	& 17.92	0.09	& 17.52	0.06	& 16.77	0.06\\
55975.46	& 19.17	0.14	& 18.01	0.09	& 17.66	0.06	& 16.72	0.06\\
55977.05	& 19.13	0.21	& 18.21	0.15	& 17.73	0.09	& 16.66	0.09\\
55979.72	& 19.43	0.17	& 18.42	0.12	& 17.87	0.07	& 16.85	0.07\\
55983.60	& 19.69	0.20	& 18.77	0.14	& 18.16	0.08	& 16.86	0.07\\
55985.64	& 19.76	0.21	& 19.04	0.17	& 18.44	0.09	& 17.00	0.07\\
55987.62 & 20.27	0.72	& 19.43	0.52	& 18.27	0.18	& 17.13	0.18\\
55988.55	& 20.02	0.27	& 19.02	0.18	& 18.67	0.11	& 17.38	0.09
\enddata

\label{tab:swiftphot}
\end{deluxetable*}

\begin{deluxetable*}{llcccll}
\footnotesize
\centering
\tablecaption{Summary of Spectroscopic Observations}
\tablecolumns{7}
\tablewidth{0pt}
\tablehead{\colhead{Date}                        &
           \colhead{MJD}                         &
           \colhead{Phase\tablenotemark{*}}      &
           \colhead{Telescope/}                  \\
           \colhead{(UT)}                        &
           \colhead{}                            &
           \colhead{(days)}                      &
           \colhead{Instrument}
}
\startdata
2012 Feb 14.79   & 55971.79  &   -3.4 & SALT/RSS  \\  
2012 Feb 16.78  & 55973.78   &   -1.4 & SALT/RSS  \\  
2012 Feb 18.80  & 55975.80   &   +0.6 & SALT/RSS  \\
2012 Feb 20.83  & 55977.83   &   +2.6 & SALT/RSS  \\
2012 Feb 21.27  & 55978.27   &   +3.1 & Keck/LRIS \\
2012 Feb 23.18  & 55980.18   &   +4.9 & Lick/Kast \\
2012 Feb 26.10     & 55983.10      &   +7.9 & HET/LRS \\
2012 Feb 27.11  & 55984.11   &   +8.9 & MMT/Blue Channel\\
2012 Mar 03.03 &  55989.03 &    +13.9 & Magellan/FIRE\\
2012 Mar 04.08     & 55990.08      &  +14.9   & HET/LRS \\
2012 Mar 15.23     & 56001.23   &  +26.0 & Keck/LRIS\\
2012 Sep 23.49     & 56193.49      & +218.3   & Keck/DEIMOS\\
2012 Oct 17.62     & 56217.62      & +242.4   & Keck/LRIS\\
2012 Oct 23.48  & 56223.48   & +248.3 & Subaru/FOCAS\\
2012 Nov 16.31  & 56247.31   & +272.1 & Magellan/IMACS\\

\enddata

\tablenotetext{*}{Phase is with respect to estimated $B$-band
maximum on Feb 18.2 (MJD 55975.2).}

\label{tab:spectra}
\end{deluxetable*}

\begin{deluxetable*}{lcccc}
\footnotesize
\centering
\tablecaption{SN\,2012ap epochs of maximum light and peak
  magnitudes}
\tablecolumns{4}
\tablewidth{0pt}
\tablehead{\colhead{Filter}    &
           \colhead{Peak Time}   &
           \colhead{Peak Time}   &
           \colhead{Peak Mag}    \\
           \colhead{}      &
           \colhead{(UT)} &
           \colhead{MJD} &
           \colhead{}
}
\startdata
$B$    & 2012 Feb 18.2 & 55975.2 $\pm 0.5$   & $17.63 \pm 0.07$ \\
$V$    & 2012 Feb 19.4 & 55976.4 $\pm 0.5$   & $16.72 \pm 0.05$ \\
$R$    & 2012 Feb 19.8 & 55976.8 $\pm 0.75$  & $16.28 \pm 0.07$ \\
$I$    & 2012 Feb 26.3 & 55983.3 $\pm 0.75$  & $15.74 \pm 0.06$
\enddata

\label{tab:lcprops}
\end{deluxetable*}

\end{document}